\documentclass[12pt,english]{revtex4-1}
\usepackage[T1]{fontenc}
\usepackage[latin9]{inputenc}
\usepackage{textcomp}
\usepackage{amsmath}
\usepackage{graphicx}
\PassOptionsToPackage{normalem}{ulem}
\usepackage{ulem}

\makeatletter

\newcommand{\lyxmathsym}[1]{\ifmmode\begingroup\def\b@ld{bold}
  \text{\ifx\math@version\b@ld\bfseries\fi#1}\endgroup\else#1\fi}

\makeatother

\usepackage{babel}
\begin{document}
\title{Effects of Dzyaloshinskii-Moriya interactions and dipole-dipole interactions
on spin waves in finite-length ferromagnetic chains}
\author{Bushra Hussain}
\email{bhussai@umich.edu}

\affiliation{Department of Natural Sciences, University of Michigan-Dearborn, Dearborn,
MI 48128, USA\bigskip{}
}
\author{Michael G. Cottam}
\email{cottam@uwo.ca}

\affiliation{Department of Physics and Astronomy, University of Western Ontario,
London, Ontario N6A 3K7, Canada\bigskip{}
}
\begin{abstract}
A spin-wave theory that includes the antisymmetric Dzyaloshinskii-Moriya
exchange interactions and long-range dipole-dipole interactions is
presented for finite-length ferromagnetic spin chains. It is found
that three different physical situations arise, depending on the direction
chosen in this geometry for the axial vector of the Dzyaloshinskii-Moriya
interactions. In some cases this leads to a tilting of the equilibrium
orientations near the ends of the chain due to interfacial effects
and with consequential effects on the spectrum of discrete dipole-exchange
spin waves. When variations are introduced for the dominant bilinear
exchange interactions at the ends of the spin chain, it is shown that
localized spin waves with spatial decay characteristics may occur.
\end{abstract}
\maketitle

\section{Introduction}

The existence of antisymmetric exchange interactions, usually referred
to as Dzyaloshinski-Moriya interactions (or DMI), has been of enduring
interest since its discovery \cite{Dzyaloshinski-1957,Moriya-1960}.
Typically, in bulk magnetic materials, the DMI may be small compared
with other symmetric bilinear exchange (such as Heisenberg exchange)
interactions. Nevertheless, the study of DMI has been important because
of its intrinsic antisymmetry with respect to the interchange of any
pair of spin sites, which has consequences for the spin-wave dispersion
relations. For example, in bulk-like materials (with no interfaces
taken into account) it has been pointed out that contributions \emph{linear}
in the wave-vector components will arise in the dispersion relations
at small wave vectors \cite{Melcher-1973,Moon-2013}, as well as the
quadratic contributions associated with the Heisenberg exchange. Until
recently, the occurrence of DMI was usually associated with either
the lack of inversion symmetry in the crystal lattices of some bulk
magnetic materials (such as MnSi and FeCoSi) or through a lowering
of symmetry at surfaces (boundaries). However, this topic has received
fresh impetus through strong enhancement effects in the DMI observable
at interfaces under suitable conditions (see, e.g., \cite{Moon-2013}.
The enhanced DMI effects may occur when a bilayer is formed between
the ferromagnetic metal and a nonmagnetic heavy metal (e.g. the Mn/W
or Fe/Ir) systems, giving rise to the so-called \emph{interfacial}
DMI (or i-DMI).

The characteristic DMI wave-vector dependence, which was mentioned
above, and its consequences for unidirectional spin-wave propagation
and/or for Brillouin light scattering (BLS), were elaborated further
by several other authors, e.g. \cite{Zakeri-2010,Cortes-2013,Kostylev-2014,Chaurasiya-2016,Tacchi-2017,Bouloussa-2020,Silvani-2021,Chen-2022}.
Some reviews covering the effects of DMI on the magnetization statics
and dynamics in ferromagnetic nanostructures are \cite{Moon-2013,Gallardo-2019,Zakeri-2020,Gallardo-2022,Shen-2022}.
In general, it is known that DMI can also produce chiral and topological
features in nanostructures \cite{Gallardo-2019,Komineas-2015,Barman-2021},
opening up new perspectives for device applications.

Many of the works referenced above are for complete magnetic thin
films or for long magnetic nanowire stripes, where there is a well-defined
wave vector in at least one direction due to translational symmetry.
Also, studies of magnonic crystals, in which there is a periodic array
of magnetic elements and hence an artificial Bloch wave vector to
the system, were also included in the above references. The cases
of DMI in `zero-dimensional' finite objects, however, deserve further
attention since these are situations where there is no well-defined
wave vector. Some examples are finite-length nanowire stripes and
flat (quasi-two-dimensional) nanorings or nanodisks with finite outer
radius. Some spin-wave calculations for the latter structures with
DMI were reported recently \cite{Flores-2020,Hussain-2022}, where
the predicted effects indicated a tilting of the spatially inhomogeneous
magnetization out of the plane of the structure in some cases (depending
on the direction of the DMI axial vector) and to modifications for
the magnetization dynamics due to DMI. Here our focus is on the dipole-exchange
spin waves in finite-length nanowire structures, with the DMI (including
end effects) taken into account.

Specifically, we will employ a one-dimensional (1D) model of a nanowire
as a finite-length chain of interacting spins. We include bilinear
(Heisenberg) exchange interactions and the long-range dipole-dipole
interactions to investigate how the dipole-exchange spin waves are
influenced by the DMI. In the absence of DMI, the spins are arranged
equally-spaced along the $x$ axis and there is a transverse applied
static magnetic field along the $z$ direction. With the DMI included,
three distinct situations can arise depending on whether the direction
of the axial vector associated with the DMI is chosen to be parallel
to the chain length (along the $x$ axis) or in one of the perpendicular
directions ($y$ or $z$ axis). All three cases are analyzed, and
it is found that the static spin orientations and the dynamics (the
mode localization and frequencies of the spin waves) are significantly
modified by the DMI due to the competing interactions. In some cases,
the termination conditions occurring at the two ends of the chain
lead to the prediction of localized spin waves that have decay characteristics
along the chain. The calculations are carried out within a microscopic
Hamiltonian operator formalism and a 1D lattice of spins.

In general, quasi-1D magnetic chain systems have been a topic of considerable
theoretical and experimental interest, following pioneering work by
Villain \cite{Villain-1959} on systems where the Ising exchange interactions
dominate and a soliton-like description for the magnetic excitation
is appropriate. Subsequently, the systems studied have involved a
broad range of different physical implementations and interactions
(see, e.g., \cite{Kim-2014,Gredig-2012,Xiang-2007,Monney-2013,Hu-2014,Diep-2015}
and an extensive review article \cite{Mikeska-2004}). In some cases
\cite{Kim-2014,Gredig-2012}, the interactions between magnetic spins
was influenced by the choice of substrate materials, e.g. in \cite{Kim-2014}
the exchange interactions producing helical order in a magnetic spin
chain originated from the indirect Ruderman-Kittel-Kasuya-Yosida (RKKY)
mechanism through an underlying electron gas in the substrate. It
is also relevant to point out that extended multilayers, such as those
formed from two magnetic materials (such as Fe/Gd) or from Permalloy
layers alternating with a nonmagnetic spacer, can often be treated
as effective 1D systems in terms of the structural variations in the
growth direction and the magnetic excitations (see \cite{Barman-2021,Gubbiotti-2019}).

The paper is organized as follows. In Sec. II the theory is presented,
taking three distinct cases according to whether the axial vector
of the DMI lies along the length of the chain or whether it is in
one of the perpendicular directions. A quantum-mechanical operator
equation-of-motion approach is employed, taking into account the competing
effects of the bilinear exchange interactions, dipole-dipole interactions,
Zeeman field energy, and the DMI and the finite chain. Since the DMI
may induce a tilting of the spins near the ends of the chain, it is
important to calculate the equilibrium orientations as a preliminary
to the spin-wave calculations. The numerical results obtained from
the theory are presented in Sec. III. Then in Sec. IV we show that
varying the termination conditions occurring at the two ends of the
chain may lead to the prediction of localized spin waves that have
decay characteristics along the chain. Finally, the conclusions are
given in Sec. V, along with generalizations to other nanostructures,
such as ferromagnetic metallic nanorings.

\section{Theory for DMI in spin chains}

\begin{figure}
\includegraphics[scale=0.7]{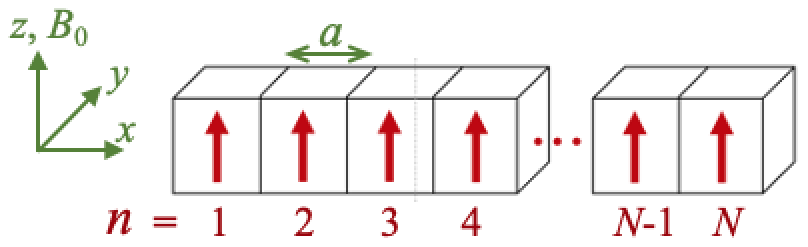}

\caption{\label{fig:chain}Geometry for a finite-length ferromagnetic chain
with $N$ interacting spins along the $x$ axis with coordinates $x_{n}=na$
where $a$ is the separation between neighbouring spins and $n=1,2,3,\cdots,N$
is a spin label. In the absence of DMI the spins are aligned along
the $z$ axis.}
\end{figure}

The assumed geometry for a finite-length ferromagnetic chain is illustrated
in Fig. \ref{fig:chain}. The system has $N$ interacting spins located
along the $x$ axis with separation $a$ between neighbours. A transverse
applied magnetic field $B_{0}$ acts in the $z$ direction, which
will stabilize the magnetic ordering in that direction when the interactions
simply consist of the short-range Heisenberg (bilinear) exchange interactions
and the long-range dipole-dipole interactions. Additionally, we consider
the effects of the antisymmetric DMI exchange contributions, which
will be shown to modify the equilibrium spin orientations by introducing
a tilting of the spins located near the ends of the chain. We are
interested here in both the static and dynamic effects of the DMI
in this dipole-exchange magnetic system.

The Hamiltonian for the ferromagnetic chain can be written in terms
of the spin operators $\mathbf{S}_{n}$ at site $n$ as

\begin{eqnarray}
\mathcal{H} & = & -\frac{1}{2}\underset{n,m}{\sum}J_{n,m}\mathbf{S}_{n}\cdot\mathbf{S}_{m}-\frac{1}{2}\underset{n,m}{\sum}\boldsymbol{\mathit{J}}_{M\mathit{n,m}}\cdot(\mathbf{S}_{n}\times\mathbf{S}_{m})\nonumber \\
 &  & +\frac{1}{2}\underset{n,m}{\sum}\underset{\alpha,\beta}{\sum}D_{n,m}^{\alpha\beta}S_{n}^{\alpha}S_{m}^{\beta}-g\mu_{B}B_{0}\underset{n}{\sum}S_{n}^{z}.
\end{eqnarray}
Here, the first two terms represent the bilinear exchange and the
DMI, respectively, with the exchange parameters $J_{n,m}$ and $\boldsymbol{J}_{Mn,m}$
being symmetric and antisymmetric with respect to the spin labels
$n$ and $m$. We note that $\boldsymbol{J}_{Mn,m}$ is an axial vector,
which may in principle be directed along any of the $x$, $y$, or
$z$ directions. The third term in the Hamiltonian is the contribution
due to the long-range dipole-dipole interactions along the chain,
where the interaction term is (with $\alpha$ and $\beta$ denoting
Cartesian componemts $x$, $y$, or $z$):
\begin{eqnarray}
D_{n,m}^{\alpha\beta} & = & (g\mu_{B})^{2}\frac{\left|\mathbf{r}_{n,m}\right|^{2}\delta_{\alpha\beta}-3r_{n,m}^{\alpha}r_{n,m}^{\beta}}{\left|\mathbf{r}_{n,m}\right|^{5}}.
\end{eqnarray}
Here $\mathbf{r}_{n,m}$ is the vector separation between sites $n$
and $m$, $g$ is the Land\'{e} factor, and $\mu_{B}$ denotes the
Bohr magneton. In the chain geometry, the only nonzero dipole-dipole
coefficients correspond to the diagonal terms $D_{n,m}^{yy}=D_{n,m}^{zz}=-\frac{1}{2}D_{n,m}^{xx}\equiv d_{n,m}$,
where $d_{n,m}=(g\mu_{B})^{2}/(|n-m|a)^{3}$ for $n\neq m$ and is
zero otherwise. The final term in Eq. (1) represents the Zeeman energy
due to the applied magnetic field $B_{0}$ in the $z$ direction.

It will be assumed that the two types of exchange terms couple only
nearest neighbours. Initially, we shall take these interactions to
have the same (bulk) values between all sites, but in a later section
end perturbations will be introduced to study spin-wave mode localization.
Specifically, we take here $J_{n,m}=J\delta_{n\pm1,m}$ and $J_{n,m}=\pm J_{M}\delta_{n\pm1,m}$
for the symmetric and antisymmetric exchange terms, where $J>0$ for
a ferromagnet but the weaker $J_{M}$ may have have either sign.

A calculation was recently given by Moon \emph{et al.} \cite{Moon-2013}
for an \emph{infinite} chain of spins with nearest-neighbour Heisenberg
exchange coupling and DMI. Here we are considering the behaviour of
\emph{finite-length} chains when DMI is present, along with an applied
field, Heisenberg exchange, and dipole-dipole interactions. Thus,
by contrast with \cite{Moon-2013}, there will be edge effects near
the ends of the chain, which may cause the spins to tilt away from
the $z$ axis (the field direction). We now explore the three different
cases for the direction of the DMI axial vector.

\subsection{DMI axis along $\boldsymbol{\hat{z}}$ direction}

We write for this case $\boldsymbol{\mathit{J}}_{M\mathit{n,m}}=\boldsymbol{\hat{z}}J_{Mn,m}$,
so the DMI term in the Hamiltonian involves the combination $\boldsymbol{\hat{z}}\cdot(\mathbf{S}_{n}\times\mathbf{S}_{m})=S_{n}^{x}S_{m}^{y}-S_{n}^{y}S_{m}^{x}.$
In order to find the equilibrium orientations of the spins along the
chain, we proceed by writing down an energy functional $\bar{E}$,
deduced from the total Hamiltonian terms in a mean field approximation.

\begin{eqnarray}
\bar{E} & = & -\frac{1}{2}\underset{n,m}{\sum}[J_{n,m}\{S_{n}^{x}S_{m}^{x}+S_{n}^{y}S_{m}^{y}+S_{n}^{z}S_{m}^{z}\}+J_{Mn,n}(S_{n}^{x}S_{m}^{y}-S_{n}^{y}S_{m}^{x})\nonumber \\
 &  & +(g\mu_{B})^{2}d_{n,m}\{2S_{n}^{x}S_{m}^{x}-S_{n}^{y}S_{m}^{y}-S_{n}^{z}S_{m}^{z}\}]-g\mu_{B}B_{0}\sum_{n}S_{n}^{z}.
\end{eqnarray}
Then, the components of the effective field $\mathbf{B}_{eff,n}$
acting on any spin $n$ are calculated from
\begin{eqnarray}
B_{eff,n}^{\alpha} & = & -\frac{1}{g\mu_{B}}\,\frac{\delta E}{\delta S_{n}^{\alpha}}\quad(\alpha=x,y,z),
\end{eqnarray}
giving rise to a set of $3N$ coupled equations ($n=1,2,\cdots N$):

\begin{eqnarray}
B_{eff,n}^{x} & = & \frac{1}{g\mu_{B}}\sum_{m}\{(J_{n,m}+d_{n,m})S_{m}^{x}+J_{Mn,m}S_{m}^{y}\},
\end{eqnarray}

\begin{eqnarray}
B_{eff,n}^{y} & = & \frac{1}{g\mu_{B}}\sum_{m}\{(J_{n,m}-\frac{1}{2}d_{n,m})S_{m}^{y}-J_{Mn,m}S_{m}^{x}\},
\end{eqnarray}

\begin{eqnarray}
B_{eff,n}^{z} & = & B_{0}+\frac{1}{g\mu_{B}}\sum_{m}(J_{n,m}S_{m}^{z}-\frac{1}{2}d_{n,m})S_{m}^{z}.
\end{eqnarray}
These equations take a simple form, where it is seen that the transverse
($x$ and $y$ components) of the effective fields do \uline{not}
couple to any longitudinal ($z$) component of a spin. This means
that the equations can all be satisfied by taking $S_{n}^{x}=S_{n}^{y}=0$
(for all $n$), and hence $S_{n}^{z}=S$. In other words, there is
no tilting effect for this particular choice of DMI axis (by contrast
with the situation for the other two choices of DMI axis later). It
is easily verified that the above solution is indeed the stable equilibrium
solution, provided $B_{0}$ is sufficiently large to overcome the
static demagnetizing effects in the $z$ direction.

Turning next to the spin-wave dynamics, we may follow steps analogous
to those in \cite{Nguyen-2005b,Lupo-2016} by transforming the spin
Hamiltonian (1) to boson operators using the Holstein-Primakoff transformation
\cite{Holstein-1940} relative to the local axes (which coincide with
the global $x$, $y$ and $z$ axes in this case). The total Hamiltonian
can be expanded as $\mathcal{H}=\mathcal{H}^{(0)}+\mathcal{H}^{(1)}+\mathcal{H}^{(2)}+\mathcal{H}^{(3)}+...$,
where $\mathcal{H}^{(s)}$ denotes a term with $s$ boson operators,
the first term is a constant and $\mathcal{H}^{(1)}$ vanishes by
symmetry. Therefore, for the linearized SWs we are concerned only
with the quadratic term, which can be written in a bilinear form as

\begin{eqnarray}
\mathcal{H}^{(2)} & = & \underset{n,m}{\sum}\{A_{n,m}a_{n}^{\dagger}a_{m}+B_{n,m}(a_{n}^{\dagger}a_{m}^{\dagger}+a_{n}a_{m})\},
\end{eqnarray}
where $A_{n,m}$ and $B_{n,m}$ may be regarded as elements of $N\times N$
matrices $\boldsymbol{A}$ and $\boldsymbol{B}$ with

\begin{eqnarray}
A_{n,m} & = & A_{n}^{0}\delta_{n,m}-SJ_{n,m}+iSJ_{Mn,m}-\frac{1}{2}Sd_{n,m},\\
A_{n}^{0} & = & B_{n,m}g\mu_{B}B_{0}+S\sum_{p=1}^{N}\{J_{n,p}-d_{n,p}\},\quad B_{n,m}=-\frac{3}{4}Sd_{n,m}.
\end{eqnarray}

The final step in determining the SW frequencies and their amplitudes
is to diagonalize $\mathcal{H}^{(2)}$ using a generalized Bogoliubov
transformation (described in \cite{Nguyen-2005b,Lupo-2016}). This
eventually gives rise to a dynamical block matrix defined by

 \begin{equation}
\left(\begin{array}{cc} \boldsymbol{A}^{(2)} & 2\boldsymbol{B}^{(2)}\\ -2\boldsymbol{B}^{(2)*} & -\tilde{\boldsymbol{A}}^{(2)} 
\end{array}\right),
\end{equation}where the tilde denotes a matrix transpose. The positive eigenvalues
of the above large matrix correspond to the total of $N$ physical
SW frequencies; there is a set of degenerate (in magnitude) frequencies
formed by the negative eigenvalues. The ``diagonalized'' form of $\mathcal{H}^{(2)}$
can be expressed as

\begin{equation}
\mathcal{H}^{(2)}=\sum_{l=1}^{N}\omega_{l}\,b_{l}^{\dagger}b_{l}, 
\end{equation}where \$\textbackslash omega\_\{l\}\$ are the discrete spin-wave
modes with integer $l=1,\,2,\,\ldots\,,N$ being a branch number,
while $b_{l}^{\dagger}$ and $b_{l}$ are the transformed (diagonalized)
boson operators for creation and annihilation of mode $l$. The eigenvectors
of the matrix in Eq. (11) provide us with the spatially-dependent
complex amplitudes \cite{Hussain-2022}, i.e. with the relative phase
information included. Numerical examples of the application of the
above results will be given in Sec. III.

\subsection{DMI axis along $\boldsymbol{\hat{x}}$ direction}

We next turn to the more interesting situation when the direction
of the DMI axis lies along the chain length $x$. The DMI term of
spin Hamiltonian now involves the combination $\boldsymbol{\hat{x}}\cdot(\mathbf{S}_{n}\times\mathbf{S}_{m})=S_{n}^{y}S_{m}^{z}-S_{n}^{z}S_{m}^{y}$.
We may follow the analogous steps as in Sec. IIB in forming an energy
functional and the effective field components on any spin. The modified
equations are

\begin{eqnarray}
B_{eff,n}^{x} & = & \frac{1}{g\mu_{B}}\sum_{m}(J_{n,m}+d_{n,m})S_{m}^{x},
\end{eqnarray}

\begin{eqnarray}
B_{eff,n}^{y} & = & \frac{1}{g\mu_{B}}\{\sum_{m}\{(J_{n,m}-\frac{1}{2}d_{n,m})S_{m}^{y}+J_{Mn,m}S_{m}^{z}\},
\end{eqnarray}

\begin{eqnarray}
B_{eff,n}^{z} & = & B_{0}+\frac{1}{g\mu_{B}}\sum_{m}\{(J_{n,m}-\frac{1}{2}d_{n,m})S_{m}^{z}-J_{Mn,m}S_{m}^{y}\}.
\end{eqnarray}
It is seen from the above that there is a coupling between the $S^{y}$
and $S^{z}$ spin components along the chain, but the expression for
the $B_{eff}^{x}$ field involves only $S^{x}$. A careful analysis
(also confirmed in the numerical calculation) shows that the spins
along the chain are tilted in the $yz$ plane away from the $z$ direction
through an angle $\theta_{n}$ for spin $n$. The effect is more pronounced
near the end of the chain; it is a consequence of the antisymmetry
of the DMI terms and missing exchange interactions at the ends.

To solve for the tilt angles, we may re-express the mean-field spin
components on the right-hand side of Eqs. (13)-(15) using $\mathbf{S}_{n}=S(0,\sin\theta_{n},\cos\theta_{n})$,
giving

\begin{eqnarray}
\textnormal{sin}\theta_{n} & = & \frac{B_{eff,n}^{y}}{\sqrt{(B_{eff,n}^{y})^{2}+(B_{eff,n}^{z})^{2}}},\quad\cos\theta_{n}=\frac{B_{eff,n}^{z}}{\sqrt{(B_{eff,n}^{y})^{2}+(B_{eff,n}^{z})^{2}}}.
\end{eqnarray}
The equilibrium orientations may then be deduced numerically solution
from Eqs. (14)-(16) through an iterative process (e.g., by analogy
with \cite{Lupo-2016,Hussain-2022}). Briefly, an initial configuration
of the angles is chosen to approximate the ground state. In practice,
we need to employ several starting configurations in order to avoid
difficulties with local minima and to find the true ground state at
$T\approx0$. For example, a configuration to approximate spin alignment
along the $z$ direction could be chosen. Next, each spin can be rotated
to be along the direction of its local effective field, giving a new
set of angles. This process can be repeated iteratively until convergence
to a self-consistent static-equilibrium configuration is achieved,
giving the required set of angles. As part of this process, it is
neccessary to check which of the configurations leads to the lowest
(global) minimum energy. When the stable spin configuration has been
determined, we may use the set of final angles to transform from the
global coordinates ($x,y,z$) to a set of local coordinates ($X,Y,Z$)
chosen such that the new $Z$ axis is along the equilibrium direction
of each spin.

Then we determine the spin-wave properties following analogous steps
to those described in the previous subsection. An important difference
is that the Holstein-Primakoff transformation to boson operators is
introduced relative to the \emph{local} spin coordinates. It is found
that Eqs. (8), (11) and (12) are still formally applicable, except
that the matrix elements are now given by

\begin{eqnarray}
A_{n,m} & = & A_{n}^{0}\delta_{n,m}-\frac{1}{2}S\{J_{n,m}[1+\cos(\theta_{n}-\theta_{m})]+J_{Mn,m}\sin(\theta_{n}-\theta_{m})\nonumber \\
 &  & +d_{n,m}[2-\cos(\theta_{n}-\theta_{m})]\},
\end{eqnarray}

\begin{eqnarray}
A_{n}^{0} & = & g\mu_{B}B_{0}\cos\theta_{n}+S\sum_{p}\{J_{n,p}\cos(\theta_{n}-\theta_{p})+J_{Mn,p}\sin(\theta_{n}-\theta_{p})\nonumber \\
 &  & -d_{n,p}\cos(\theta_{n}-\theta_{p})\},
\end{eqnarray}

\begin{eqnarray}
B_{n,m} & = & -\frac{1}{4}S\{J_{n,m}[1-\cos(\theta_{n}-\theta_{m})]-J_{Mn,m}\sin(\theta_{n}-\theta_{m})\nonumber \\
 &  & +d_{n,m}[2+\cos(\theta_{n}-\theta_{m})]\}.
\end{eqnarray}

\subsection{DMI axis along $\boldsymbol{\hat{y}}$ direction}

The final case to consider is when the DMI axial vector is along the
$y$ direction, for which $\boldsymbol{\hat{y}}\cdot(\mathbf{S}_{n}\times\mathbf{S}_{m})=S_{n}^{z}S_{m}^{x}-S_{n}^{x}S_{m}^{z}$.
It is straightforward to show that the coupled equations for the components
of the effective mean fields become

\begin{eqnarray}
B_{eff,n}^{x} & = & \frac{1}{g\mu_{B}}\{\sum_{m}(J_{n,m}+d_{n,m})S_{m}^{x}-\sum_{m}J_{Mn,m}S_{m}^{z}\},
\end{eqnarray}

\begin{eqnarray}
B_{eff,n}^{y} & = & \frac{1}{g\mu_{B}}\sum_{m}(J_{n,m}-\frac{1}{2}d_{n,m})S_{m}^{y},
\end{eqnarray}

\begin{eqnarray}
B_{eff,n}^{z} & = & B_{0}+\frac{1}{g\mu_{B}}\{\sum_{m}(J_{n,m}-\frac{1}{2}d_{n,m})S_{m}^{z}+J_{Mn,m}S_{m}^{x}\}.
\end{eqnarray}
Here it is seen that that there is a coupling between the $S^{x}$
and $S^{z}$ spin components along the chain, while the expression
for the $B_{eff}^{y}$ field involves only $S^{y}$. By analogy with
the previous subsection, if follows that the spins along the chain
are now tilted in the $xz$ plane away from the $z$ direction through
an angle denoted as $\phi_{n}$. Instead of Eq. (16), we now have
$\mathbf{S}_{n}=S(\sin\phi_{n},0,\cos\phi_{n})$, with

\begin{eqnarray}
\textnormal{sin}\phi_{n} & = & \frac{B_{eff,n}^{x}}{\sqrt{(B_{eff,n}^{x})^{2}+(B_{eff,n}^{z})^{2}}},\quad\cos\phi_{n}=\frac{B_{eff,n}^{z}}{\sqrt{(B_{eff,n}^{x})^{2}+(B_{eff,n}^{z})^{2}}}.
\end{eqnarray}
In a numerical application the equilibrium tilt angles may be found
through an iterative process, as described earlier.

The calculations for the spin-wave excitations then proceeds in an
analogous fashion to the previous subsection. The coefficients of
the quadratic Hamiltonian involve the tilt angles $\phi_{n}$ and
have the modified form

\begin{eqnarray}
A_{n,m} & = & A_{n}^{0}\delta_{n,m}-\frac{1}{2}S\{J_{n,m}[1+\cos(\phi_{n}-\phi_{m})]-J_{Mn,m}\sin(\phi_{n}-\phi_{m})\nonumber \\
 &  & +d_{n,m}[(2\cos\theta_{n}\cos\theta_{m}-\sin\theta_{n}\sin\theta_{m}-1)\},
\end{eqnarray}

\begin{eqnarray}
A_{n}^{0} & = & g\mu_{B}B_{0}\cos\phi_{n}+S\sum_{p}\{J_{n,p}\cos(\phi_{n}-\phi_{p})-J_{Mn,p}\sin(\phi_{n}-\phi_{p})\nonumber \\
 &  & +d_{n,p}(2\sin\phi_{n}\sin\phi_{p}-\cos\phi_{n}\cos\phi_{p})\},
\end{eqnarray}

\begin{eqnarray}
B_{n,m} & = & \frac{1}{4}S\{J_{n,m}[1-\cos(\phi_{n}-\phi_{m})]+J_{Mn,m}\sin(\phi_{n}-\phi_{m})\nonumber \\
 &  & -d_{n,m}(2\cos\phi_{n}\cos\phi_{m}+\sin\phi_{n}\sin\phi_{m}+1)\}.
\end{eqnarray}

\section{Numerical results}

In this section we present some numerical applications for the quantized
spin-wave modes of finite chains, using the formal results for the
three directions of DMI axial vector obtained in Sec. II. Examples
will be given for several chain lengths ($N=15,$ 25, and 60), for
DMI values ranging from $J_{M}/J=0$ up to 0.35, and for dipole-dipole
strengths (relative to the bilinear exchange) such that $d=(g\mu_{B})^{2}/Ja{}^{3}=0$,
0.01, and 0.02. For convenience, we choose spin quantum number $S=1$
and an applied magnetic field such that $g\mu_{B}B_{0}/J=0.25.$

\begin{figure}
\includegraphics[scale=0.6]{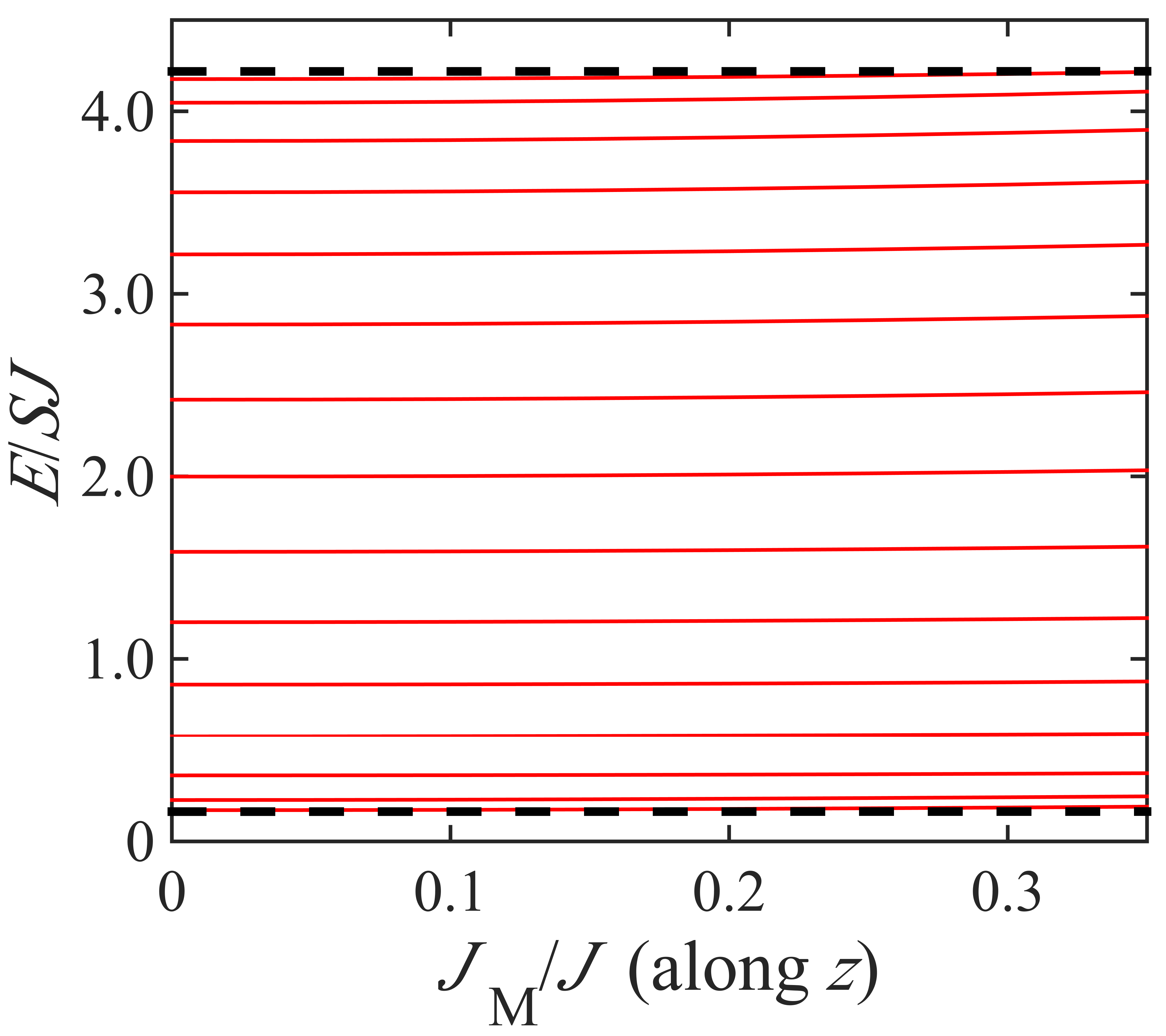}

\caption{\label{fig:2} Plot of the spin-wave energy $E$ (in terms of the
dimensionless $E/SJ$) versus relative DMI strength $J_{M}/J$ for
a fixed $N=15$ showing all 15 branches (solid red lines). The DMI
axis has been taken along the $z$ direction. The black dashed lines
indicate the upper and lower boundaries of the bulk-mode region. See
the text for other parameter values.}

\end{figure}

First, in Fig. \ref{fig:2} a plot is given of the spin-wave energy
$E$ (in dimensionless units $E/SJ$) versus the DMI strength $J_{M}/J$
for a relatively short chain with $N=15$ and a fixed value $d=$
0.02 of the dipolar strength. In this case (which is for the DMI along
$z$) the variation with DMI strength on the individual modes is evident,
giving a downwards shift for the lower modes and an upwards shift
for the higher modes. This difference of behaviour is attributable
to the fact that the amplitude oscillations for the lower modes are
all in phase, whereas those for the higher modes are 180$\lyxmathsym{\textdegree}$
out of phase for any site with respect to its nearest neighbours (see
later for discussion of the amplitudes). All the spin-wave modes in
this example are bulk modes with an oscillatory amplitude profile;
the upper and upper boundaries of the bulk mode quasi-continuum are
indicated by the black dashed lines in Fig. \ref{fig:2}.

\begin{figure}
\includegraphics[scale=0.29]{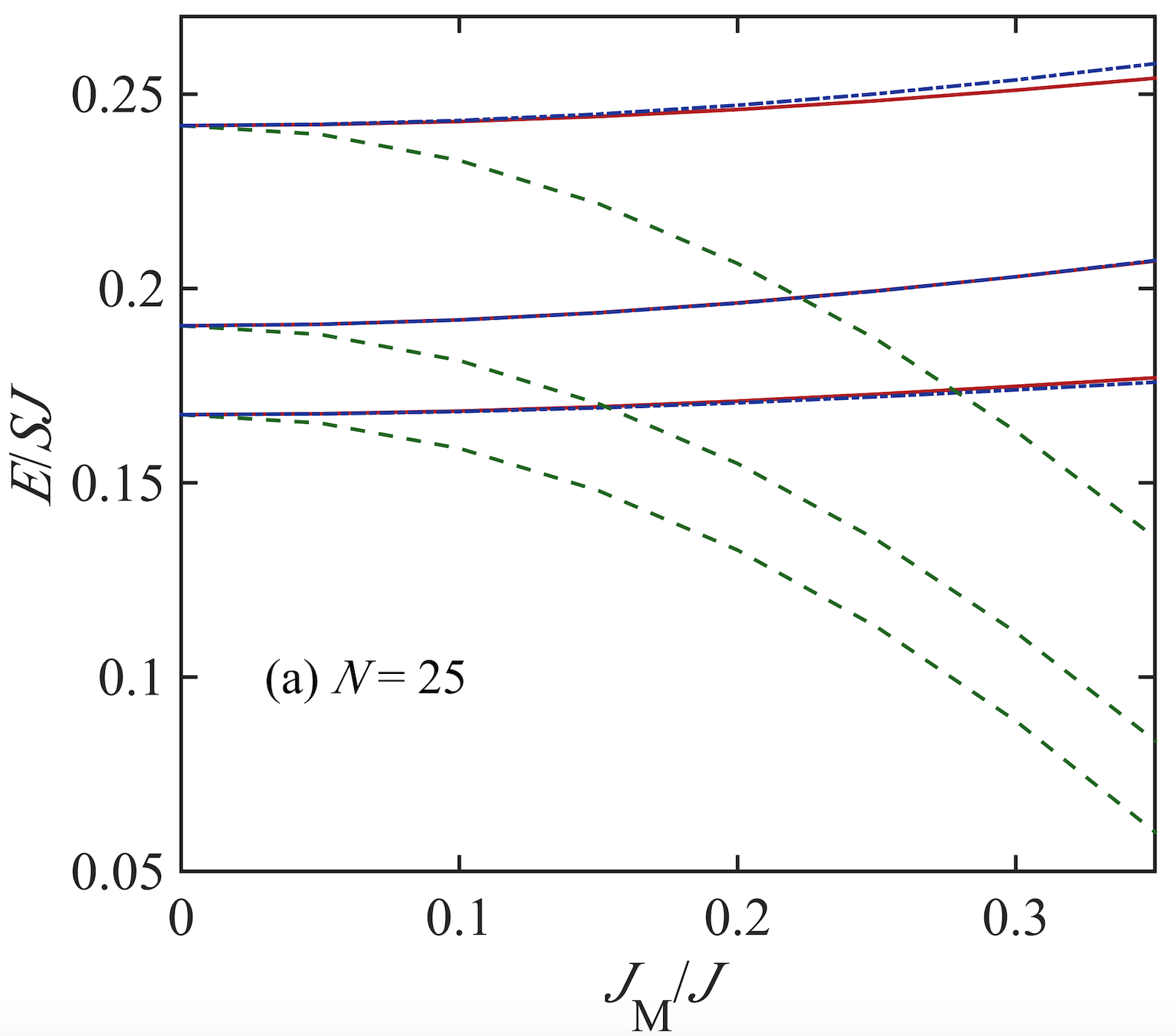}\includegraphics[scale=0.29]{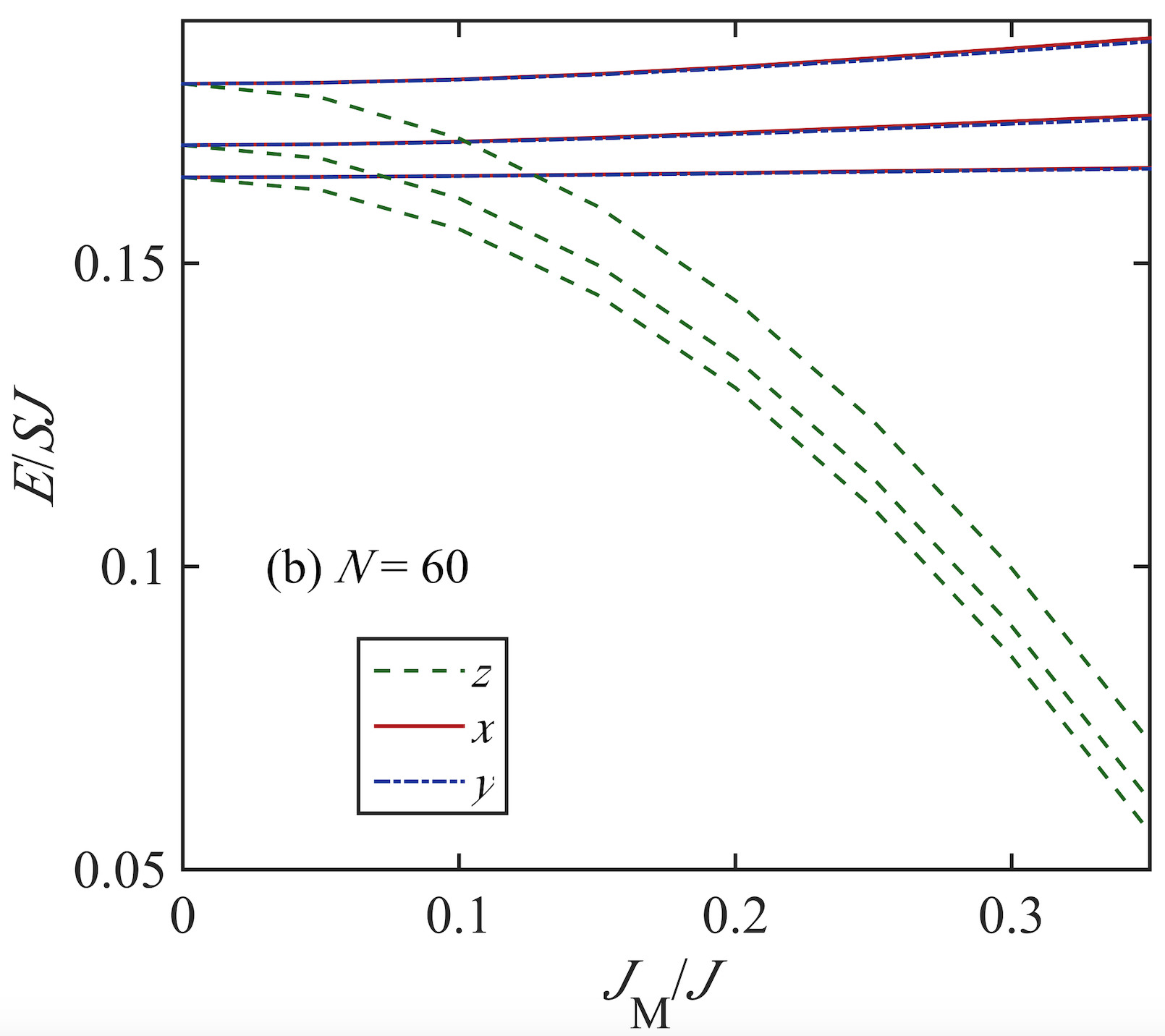}

\caption{\label{fig:3} Plot of the spin-wave energy $E$ (in terms of the
dimensionless $E/SJ$) versus relative DMI strength $J_{M}/J$ for
(a) $N=25$ and (b) $N=60$ showing only the lowest three branches
in each case. The three different types of lines (see the inset) correspond
the the DMI axis along $x$, $y$, and $z$. See the text foe other
parameter values.}
\end{figure}

\begin{figure}
\includegraphics[scale=0.4]{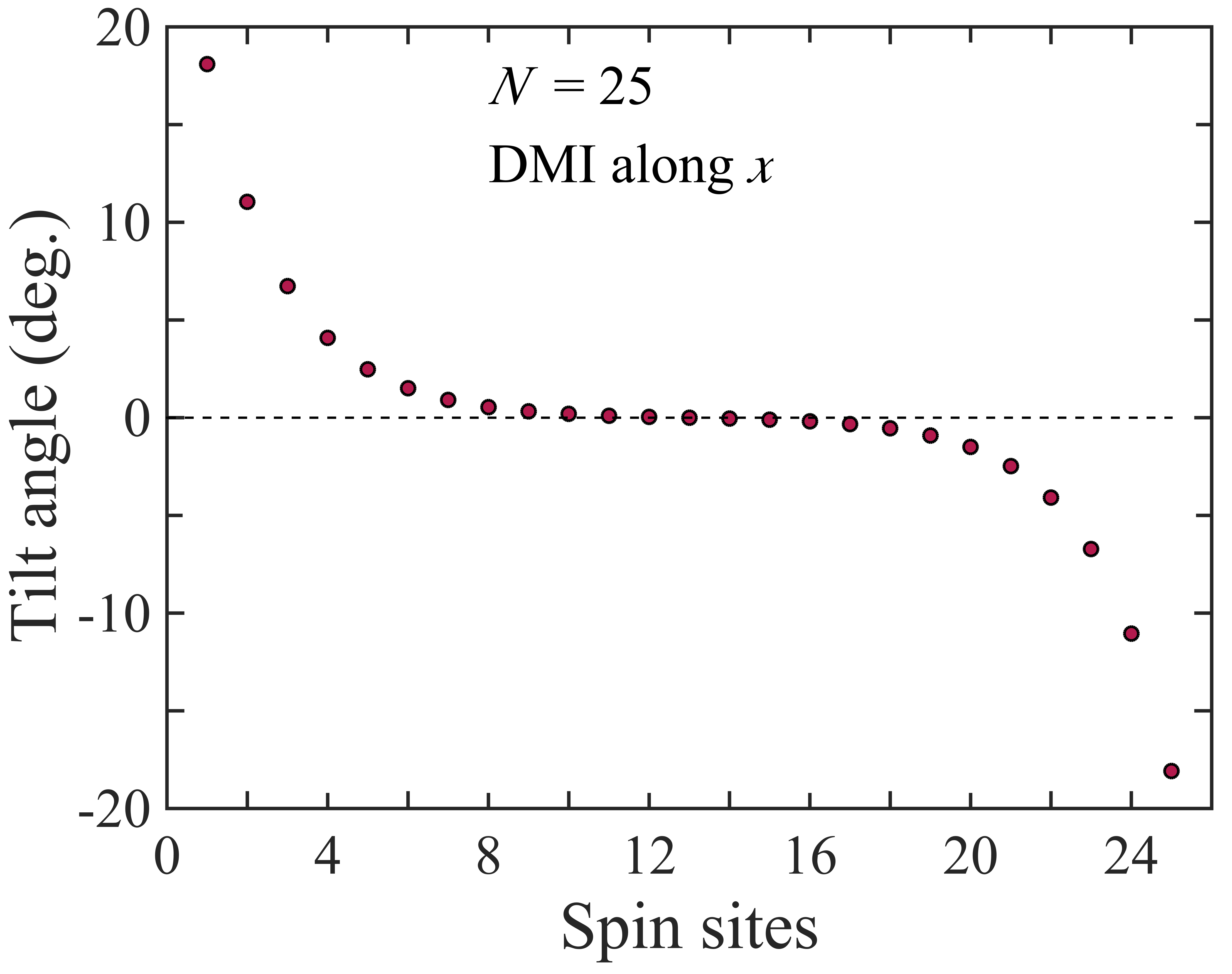}

\caption{\label{fig:4-1}Plot of the tilt angles $\theta_{n}$ (in degrees)
versus the spin number $n$ ($=1,2,\cdots,N$) choosing a chain with
$N=25$ as in Fig. \ref{fig:3}a and $J_{M}/J=0.2$.}

\end{figure}

Next, in Fig. \ref{fig:3} we show some comparisons for the bulk-mode
energies when the DMI axis is along the different principal directions
($x$, $y$ or $z$). Here the plots versus $J_{M}/J$ are shown for
just the lowest three branches for longer chain lengths corresponding
to (a) $N=25$ and (b) $N=60$. It is seen that, while the curves
for the DMI along $z$ shift downwards with increasing $J_{M}/J$,
as mentioned earlier, the curves for for the DMI along $x$ and $y$
shift slightly upwards. This difference in behaviour is presumed to
be related to tilting of the end spins in the $x$ and $y$ cases.
We illustrate in Fig. \ref{fig:4-1} some calculated values for the
tilt angle $\theta_{n}$ for different spin site number $n$ when
$N=25$ (as in Fig. \ref{fig:3}a) and $J_{M}/J=0.2$. It can be seen
that $\theta_{n}$ is largest in magnitude at the ends of the chain
and the mode profile is antisymmetric with respect to the ends.

\begin{figure}
\includegraphics[scale=0.48]{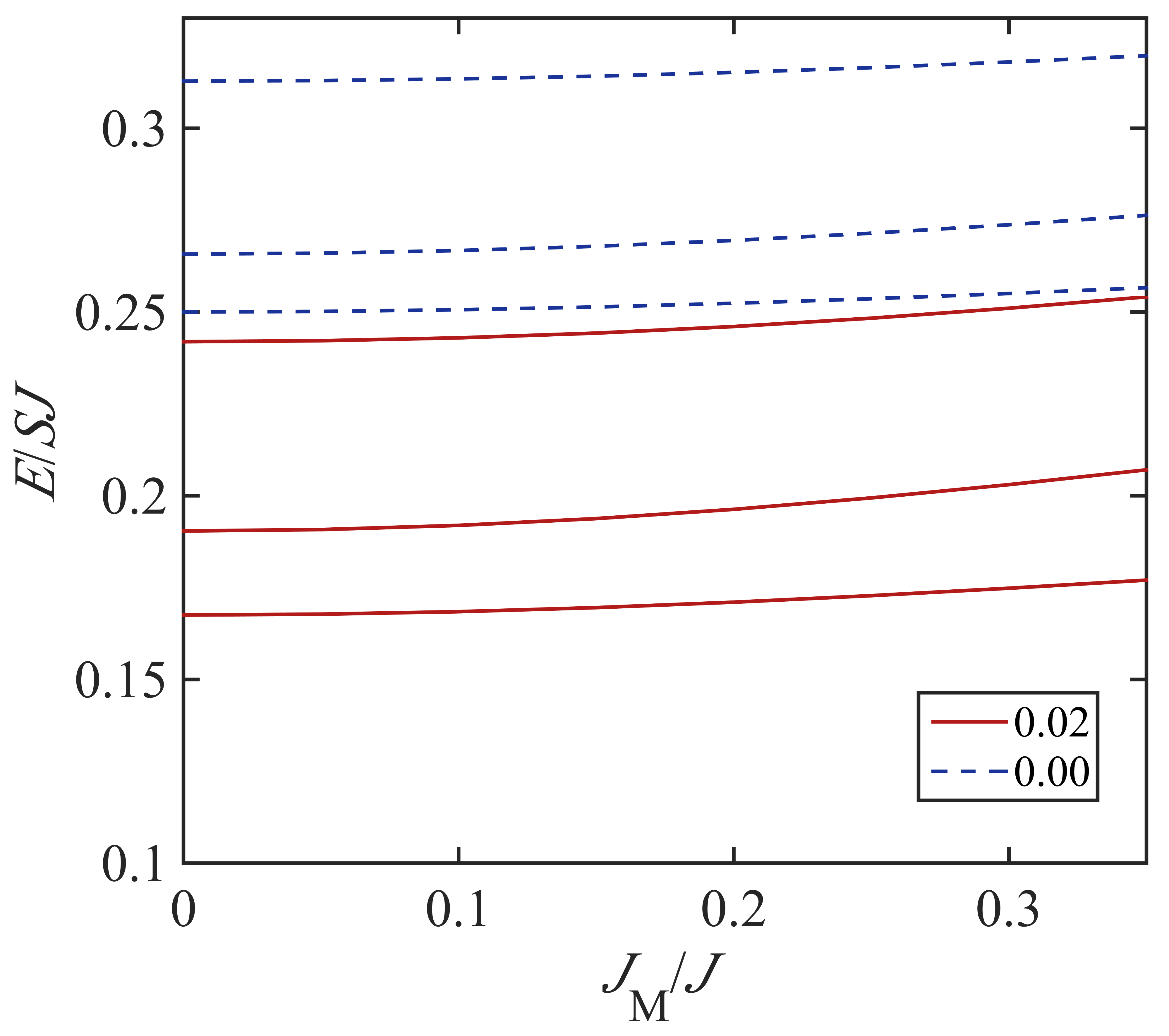}\caption{\label{fig:5-1} Plot of the spin-wave energy $E$ (in terms of the
dimensionless $E/SJ$) versus relative DMI strength $J_{M}/J$, showing
curves for three different dipole-dipole strengths $d=0.00$, 0.01,
and 0.02. We depict only the three lowest spin-wave modes for a chain
with $N=25$ for the case of DMI axis along $x$.}

\end{figure}

In Fig. \ref{fig:5-1} we illustrate the role of the dipole-dipole
interactions on the spin waves in this chain geometry. The plots in
each case, which correspond to the DMI axis along $x$, are shown
for dipole strength $d=0.0$ and 0.02, as indicated. The results are
for just the lowest three modes in a chain with $N=25$. Note that
$E$ for each spin-wave mode decreases, as expected, when the strength
$d$ is increased.

\begin{figure}
\includegraphics[scale=0.4]{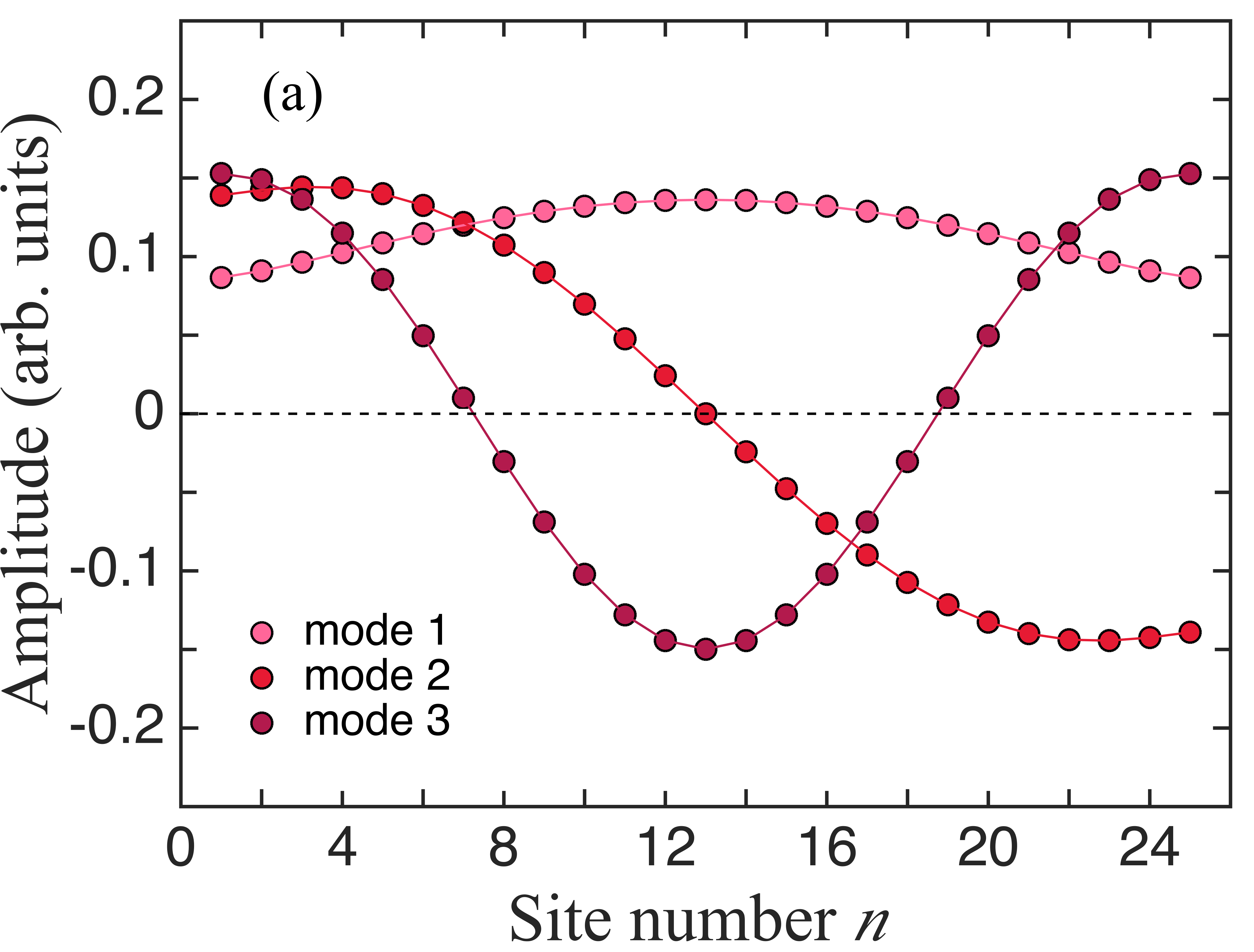}\includegraphics[scale=0.4]{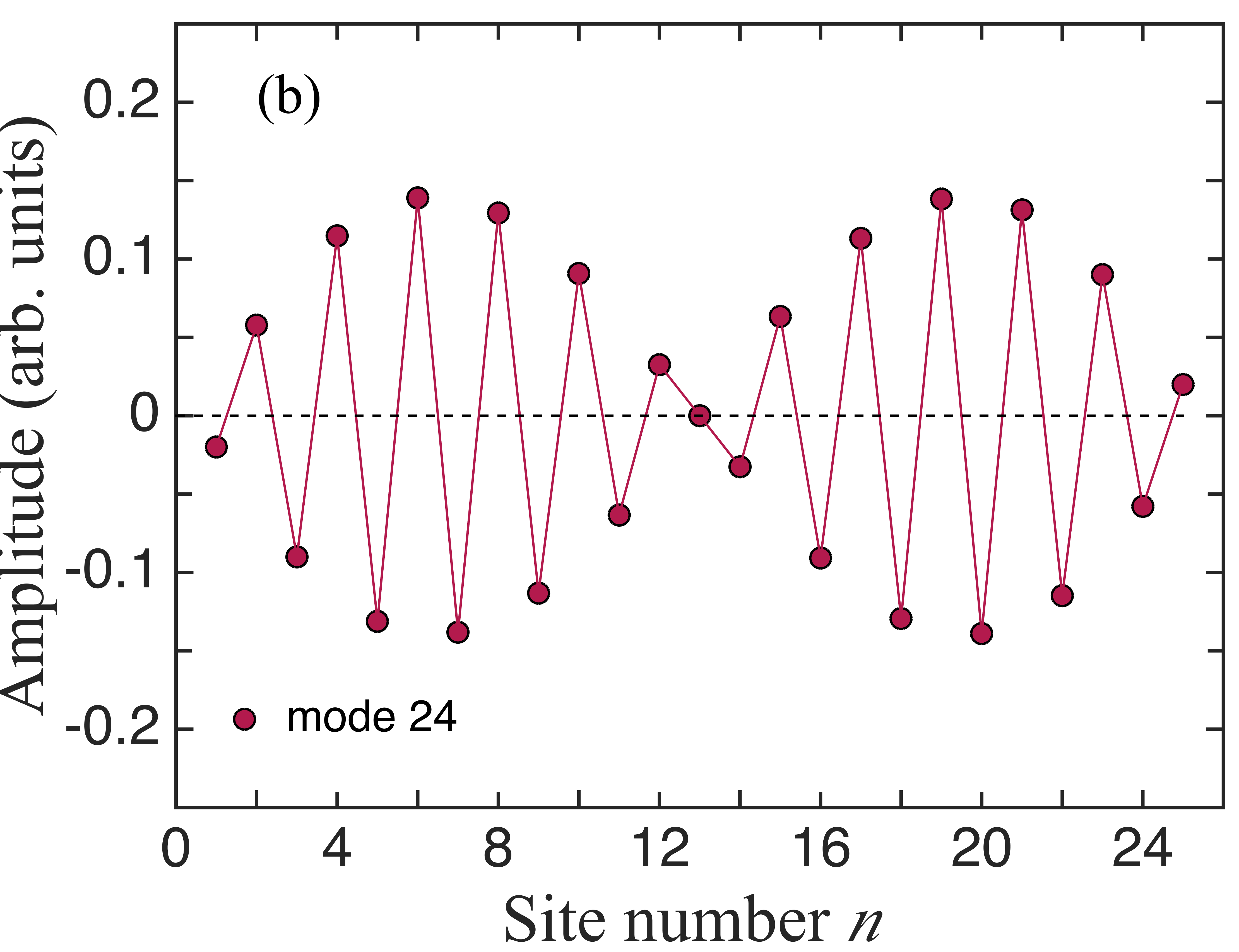}

\caption{\label{fig:6-1} Plots to show the spin-wave amplitudes (in arbitrary
units) versus the spin number $n$ ($=1,2,\cdots,N$) choosing a chain
with $N=25$. The modes 1-3 are colour-coded as indicated in the inset
to (a) and mode 24 is shown in (b). The DMI strength is $J_{M}/J$
= 0.2 along $x$ and the dipole-dipole strength is $d$ = 0.02. See
the text for other parameters.}

\end{figure}

To conclude this section we present in Fig. \ref{fig:6-1} some plots
of the spin-wave amplitude (with the relative phase included) versus
the site number $n$ for the spins along the chain. We employ the
eigenvector formalism described in Sec. II. In panel (a) we show the
profiles for the lowest three modes. Mode 1 is the analogue of the
uniform mode (but it shows some variation due to end effects), while
modes 2 and 3 are wave-like in character. For the higher energy modes,
like for mode 24 shown in panel (b), the complex amplitudes change
sign (have a 180$\lyxmathsym{\textdegree}$ change of phase) in going
from any one site to its neighbour. This was mentioned earlier for
the interpretation of Fig. \ref{fig:2}.

\section{Mode localization effects}

Up to now, all the spin-wave modes have had bulk-like propagation
properties. We now examine the possibility of modifications to the
end parameters of the chain in order to study possibilities for the
occurrence of localized spin-wave modes. As a simple assumption, we
will take the dominant nearest-neighbour symmetric bilinear exchange
$J_{n,m}$ to be different at both ends of the chain. Therefore, we
assume $J_{1,2}=J_{2,1}=J_{N-1,N}=J_{N,N-1}\equiv J_{S}$, whereas
for all other nearest neighbours it is equal to the bulk value $J.$
The value of $J_{S}$ will typically depend on overlap integrals between
wave functions at the pair of sites, and so may be modified compared
with spins in the middle of the chain. In principle, $J_{S}$ can
be either greater than or less than $J$.

The above property for $J_{S}$ can straightforwardly be accommodated
into the theory given in Sec. II. In fact, all the formal expressions
derived there still apply provided $J_{S}>0$. It is necessary only
to use the generalized assumption for $J_{n,m}$ in the numerical
calculations for the equilibrium tilt angles and for finding the eigenvalues
and eigenvectors of the dynaical matrix in Eq. (11).

It is interesting to note that there is a special case in which the
solution for the dynamical matrix can be carried out analytically.
This occurs in the case of the DMI axial vector along $z$ (so the
tilt angles are zero) when the dipole-dipole interactions are absent
($d=0$) \emph{and} the chain is semi-infinite ($N\rightarrow\infty$).
Then matrix $\boldsymbol{B}$ vanishes and matrix $\boldsymbol{A}$
simplifies, leaving the dynamical matrix expressible in a tridiagonal
form as

\begin{equation}
SJ\left(\begin{array}{cccccc}
c_{0} & -\gamma+i\alpha & 0 & 0 & 0 & \cdots\\
-\gamma-i\alpha & c' & -1+i\alpha & 0 & 0 & \cdots\\
0 & -1-i\alpha & c & -1+i\alpha & 0 & \cdots\\
0 & 0 & -1-i\alpha & c & -1+i\alpha & \cdots\\
0 & 0 & 0 & -1-i\alpha & c & \cdots\\
\vdots & \vdots & \vdots & \vdots & \vdots & \ddots
\end{array}\right),
\end{equation}
where we define the ratios $\gamma=J_{S}/J$ and $\alpha=J_{M}/J$,
while $c_{0}=\gamma+(g\mu_{B}B_{0}/SJ)$, $c'=1+\gamma+(g\mu_{B}B_{0}/SJ)$,
and $c=2+(g\mu_{B}B_{0}/SJ)$. It follows that the perturbing effects
of the end of the chain are confined to the top left $2\times2$ block
of this matrix, while elsewhere the matrix elements are constant along
the main diagonal and the diagonal lines above and below. This is
of significance here, because it is well known that the determinant
of tridiagonal matrices with these properties can be found analytically
(see, e.g., \cite{Dewames-1969,Wolfram-1972,Cottam-1976} for semi-infinite
Heisenberg ferromagnets). Hence the eigenvaluesof the dynamical matrix,
which are just the spin-wave energies, can be deduced. The result
from following an analogous approach here is that a localized surface
spin wave can exist only when $J_{S}/J$ exceeds a threshold value
of 4/3, or approximately 1.33, in which case the surface-mode energy
lies \emph{above} the top of the bulk region. The derivation is outlined
in the Appendix.

\begin{figure}
\includegraphics[scale=0.5]{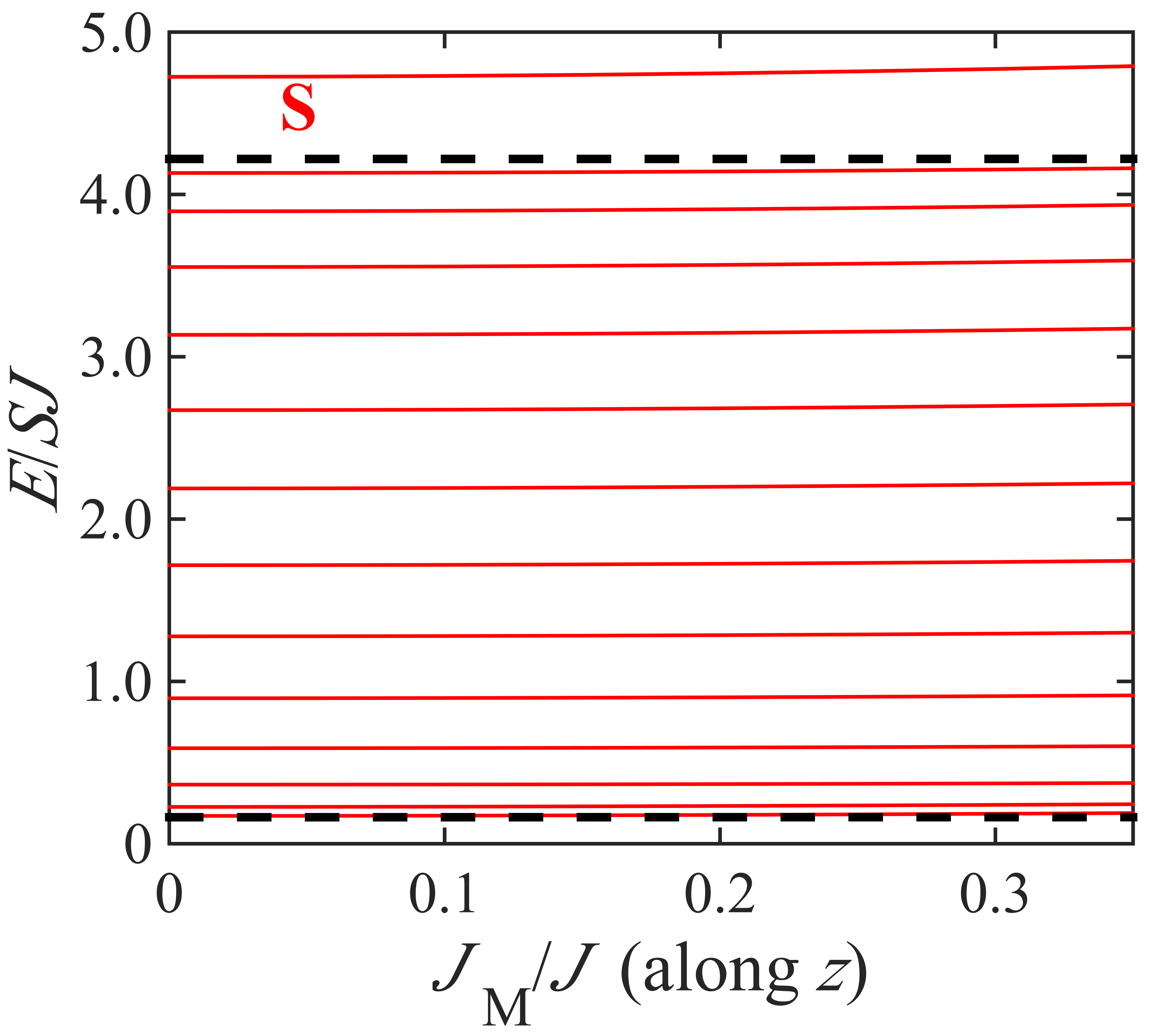}

\caption{\label{fig:7-1} Plot of the spin-wave energy $E$ (in terms of the
dimensionless $E/SJ$) versus relative DMI strength $J_{M}/J$ for
fixed $N=15$ showing all the spin-wave branches when $J_{S}/J=1.8$
taking the DMI axial vector along $z$. The black dashed lines indicate
the upper and lower boundaries of the bulk-mode region. Other parameters
are as in Fig. \ref{fig:2}.}

\end{figure}

We now present some numerical results obtained, as described earlier,
for the general case, i.e, with dipole-dipole interactions included
and with $N$ taking a finite value. In Fig. \ref{fig:7-1} a plot
is shown for the spin-wave energies versus $J_{M}/J$ for the DMI
axial vector along $z$ when $J_{S}/J$ = 1.8 and $N$ = 15. It is
seen that there are thirteen modes within the bulk-mode region (between
the black dashed lines) and two almost-degenerate modes labelled S
above the top of the bulk region. The latter are the surface modes
(one at each end of the chain). This plot may be compared with Fig.
\ref{fig:2} for $J_{S}/J$ = 1.0, where all fifteen of the spin-wave
modes are bulk-like.

\begin{figure}
\includegraphics[scale=0.5]{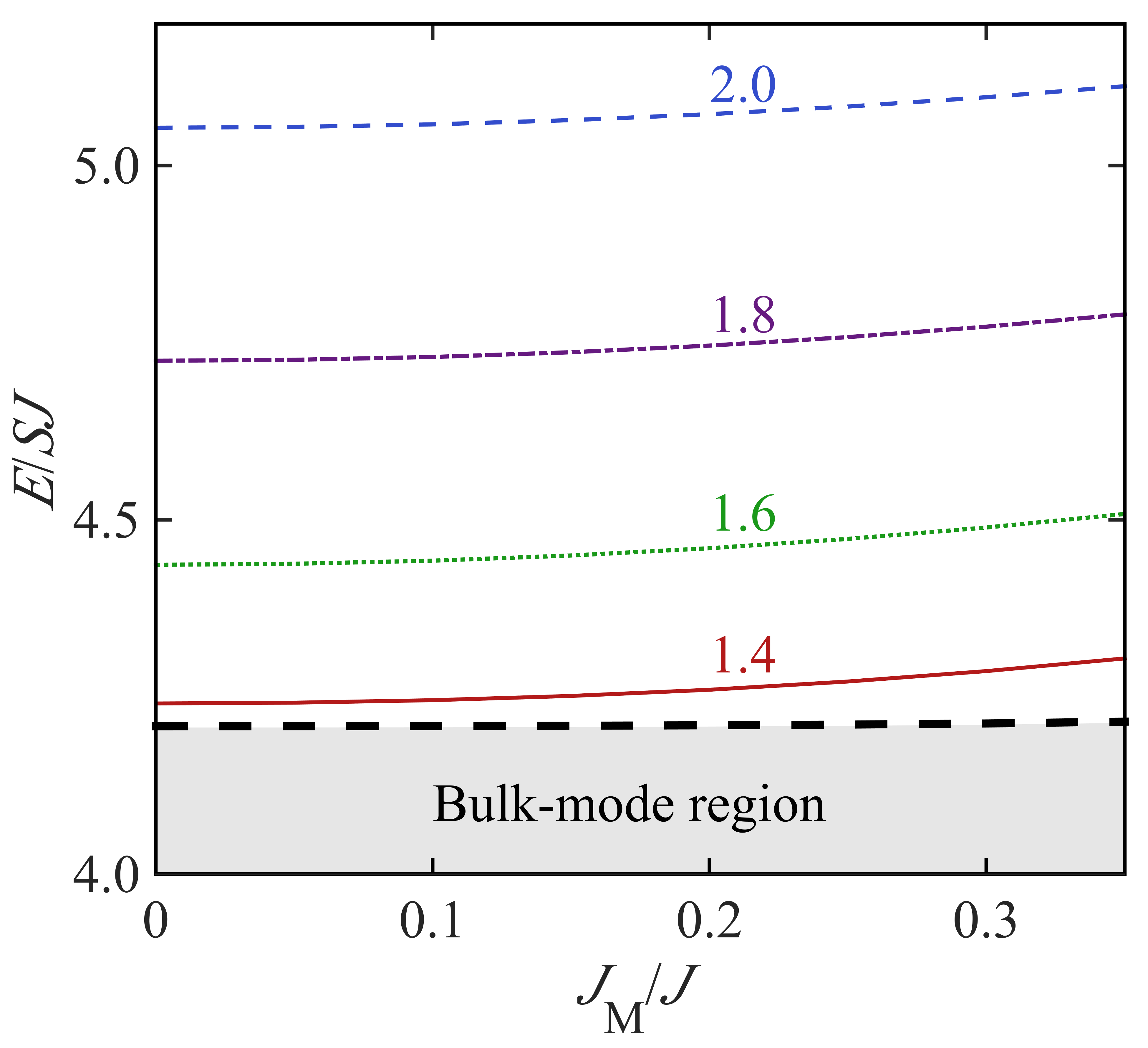}

\caption{\label{fig:8-1} Plot of spin-wave energy $E$ (in terms of the dimensionless
$E/SJ$) versus relative DMI strength $J_{M}/J$ for fixed $N=60$
showing only the surface spin-wave branches for several values of
$J_{S}/J$ (= 1.4, 1.6, 1.8 and 2), as labelled. Here the DMI axial
vector is along the $x$ axis. The black dashed line indicate the
upper boundary of the bulk-mode region.}

\end{figure}

In Fig. \ref{fig:8-1} we show the several surface spin energies for
different choices of $J_{S}/J$ (as labelled) when plotted versus
$J_{M}/J$. Here $N$ = 60 and the DMI axial vector is along $x$.
The surface modes are found occur at higher energies as $J_{S}/J$
is increased, starting here at 1.4 which is just above the threshold
value estimated earlier.

\begin{figure}
\includegraphics[scale=0.25]{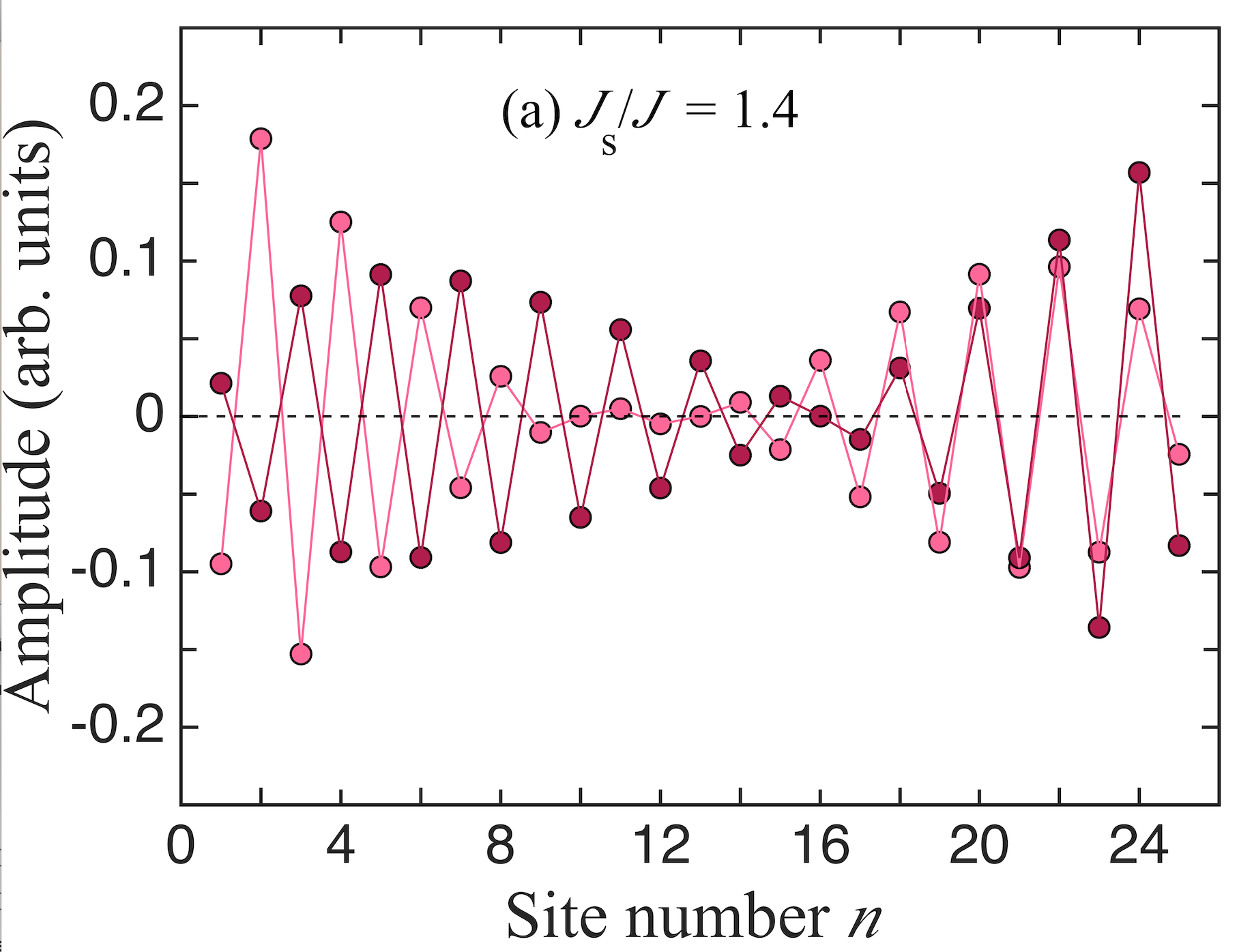}\includegraphics[scale=0.25]{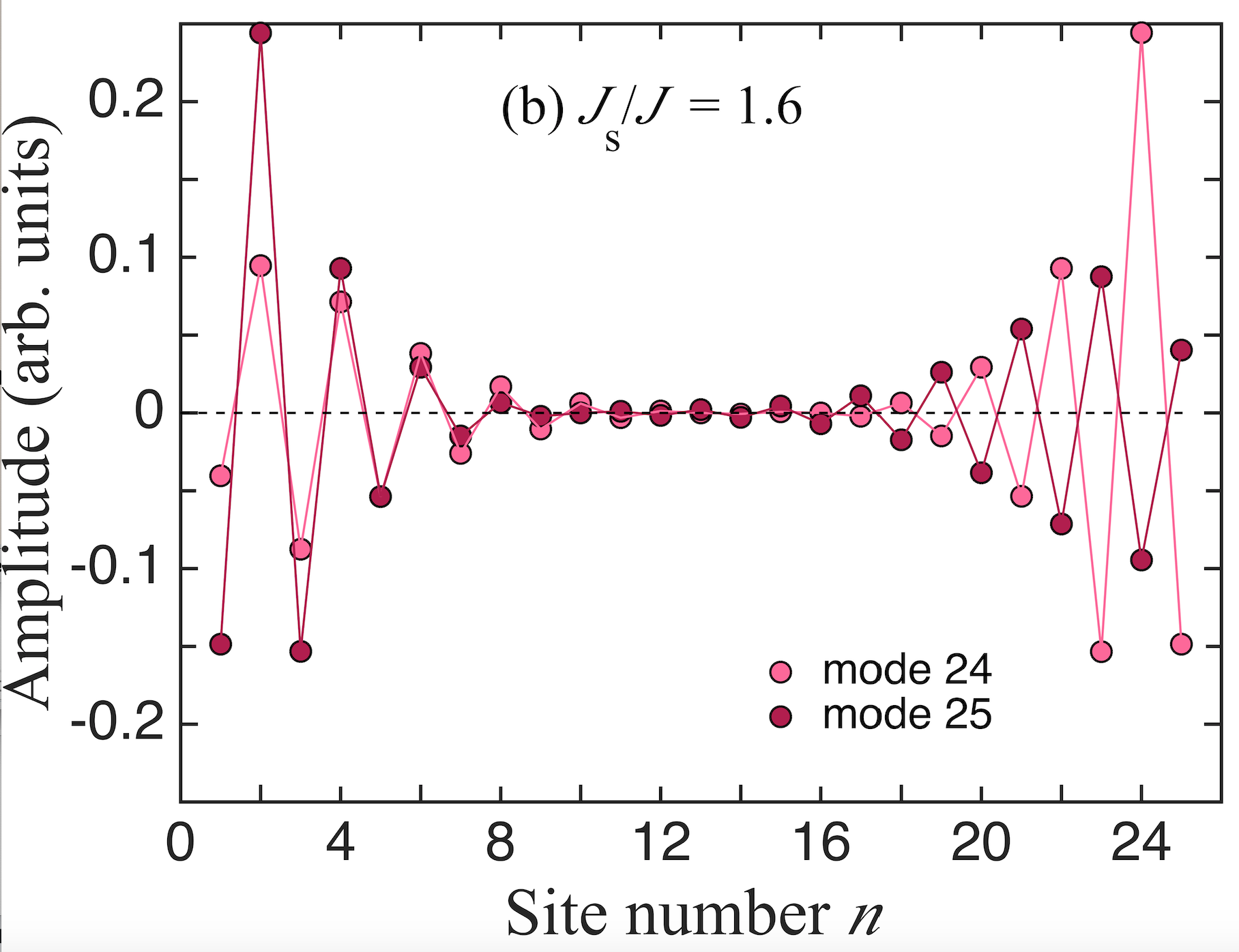}

\caption{\label{fig:9} Plots to show the surface spin-wave amplitudes (in
arbitrary units) when $J_{S}/J\protect\neq1$ versus the spin number
$n$ ($1,2,\cdots,N$) choosing a chain with $N=25$. The data are
shown for (a) $J_{S}/J=$ 1.4 and (b) $J_{S}/J=$ 1.6. We have taken
$d=$ 0.02, $J_{M}/J=$ 0.2, and the DMI axial vector along $z$.}
\end{figure}

Finally, in Fig. \ref{fig:9} we show amplitude plots for the surface
spin waves at two different values of $J_{S}/J$ (1.4 in panel a and
1.6 in panel b) versus the spin number $n$ choosing a chain with
$N=25$. Modes 24 and 25 are shown, representing the two almost degenerate
surface spin waves. It is seen that the amplitude profiles have decay
characteristics away from the ends of the chain, by contrast with
the behaviour in Fig. \ref{fig:6-1}(b) when $J_{S}/J$ = 1.0. Also,
the spatial decay is more rapid in Fig. \ref{fig:9} for the modes
with the higher value of $J_{S}/J$ which correspond to the higher
energy.

\smallskip{}

\section{Conclusions}

In this paper we have presented theoretical studies for the magnetization
dynamics for finite-length spin chains in the presence of DMI. We
included bilinear exchange interactions and the long-range dipole-dipole
interactions within a microscopic Hamiltonian operator formalism to
investigate how the dipole-exchange spin waves are influenced by the
DMI. We found that, when the DMI is included, three physically distinct
situations arise depending on whether the direction of the axial vector
associated with the DMI is chosen to be parallel to the chain length
(along the $x$ axis) or in one of the perpendicular directions ($y$
or $z$ axis). All three cases were analyzed, and it was found that
the static spin orientations and the dynamics (the mode localization
and frequencies of the spin waves) are significantly modified by the
DMI due to the competing interactions, which include the long-range
dipole-dipole interactions. In some cases, the termination conditions
occurring at the two ends of the chain lead to the prediction of localized
spin waves (occurring above the bulk band) that have decay characteristics
along the chain.

Although the DMI calculations presented here have been presented in
terms of a finite linear-chain geometry, they are of wider applicability.
For example, with only minor modifications to allow for additional
contributions to the static dipolar fields, the results obtained here
can be applied to finite-width nanowire stripes when the spin waves
are excited with zero wave vector along the nanowire length. This
work also provides a stepping stone to further DMI studies in which
generalizations are made to other nanostructures, such as nanorings
with different directions chosen for the DMI axial vector.

\section*{Appendix}

Here we outline the steps involved in deducing the mode localization
condition and mode energy from the tridiagonal matrix in Eq. (25).
It is seen that the $2\times2$ matrix $\boldsymbol{\Delta}$ representing
the perturbation has the matrix elements $\Delta_{1,1}=c_{0}-c=\gamma-2$,
$\Delta_{2,2}=c'-c=\gamma-1$, and $\Delta_{1,2}=\Delta_{2,1}=-\gamma+1$.
We may then employ the general result expressed in Eq. (60) of \cite{Cottam-1976}
that any localized surface spin wave (having decaying amplitude away
from the surface of the semi-infinite structure) has energy $E_{S}$
where

\begin{equation}
E_{S}=c-(x+x^{-1}),
\end{equation}
where $|x|<1$ as a localization condition and the factor $x$ satisfies
the cubic equation

\begin{eqnarray}
h(x) & \equiv & 1+x(\Delta_{1,1}+\Delta_{2,2})+x^{2}(2\Delta_{1,2}+\Delta_{1,1}\Delta_{2,2}-\Delta_{1,2}^{2})+x^{3}\Delta_{2,2}\nonumber \\
 & = & 1+x(2\gamma-3)-3x^{2}(\gamma-1)+x^{3}(\gamma-1)=0.
\end{eqnarray}
A careful analysis shows that a physical solution for $x$ exists
only if $\gamma>4/3$, which is the condition stated in Sec. IV. The
mode energy, which is obtained using Eq. (28), then satisfies $E_{S}>c+2,$
which means that it occurs \emph{above} the top of the bulk band.
\begin{acknowledgments}
We gratefully acknowledge support from the Natural Sciences and Engineering
Research Council (NSERC) of Canada, through Discovery Grant RGPIN-2017-04429. 
\end{acknowledgments}

\bibliographystyle{unsrt}
\bibliography{References}

\end{document}